\documentclass[conference]{IEEEtran}
\IEEEoverridecommandlockouts

\usepackage{cite}
\usepackage{amsmath,amssymb,amsfonts}
\usepackage{algorithm}
\usepackage{algorithmic}
\usepackage{graphicx}
\usepackage{url}
\usepackage{textcomp}
\usepackage{xcolor}
\usepackage{booktabs}
\usepackage{multirow}
\usepackage{siunitx}
\usepackage{placeins}
\usepackage{array}
\newcolumntype{L}[1]{>{\raggedright\arraybackslash}p{#1}}
\def\BibTeX{{\rm B\kern-.05em{\sc i\kern-.025em b}\kern-.08em
    T\kern-.1667em\lower.7ex\hbox{E}\kern-.125emX}}

\begin{document}

\title{Compliance-Scored Best-of-$N$ Guardrail Orchestration for Multimodal Document Generation in Payments Dispute Defense}

\author{
\IEEEauthorblockN{Nataraj Agaram Sundar \quad Tejas Morabia}
\IEEEauthorblockA{\textit{eBay Inc.}\\
San Jose, CA, USA\\}
}

\maketitle

\begin{abstract}
High-stakes enterprise document generation (e.g., financial dispute narratives, compliance notices, audit summaries) demands \emph{schema correctness}, \emph{policy compliance}, and \emph{low-latency} operation at scale. Prior to a unified guardrail layer, production systems often stitched together separate PII redaction, content moderation, and format validation steps, leading to fragmented logic, slower request paths, and higher operational cost.

We present a \emph{guardrail orchestration layer} for \emph{text and image} inputs that couples multi-candidate generation with an explicit \emph{compliance score} used for early exit. The framework runs configurable parallel ``heads,'' scores candidates against weighted guardrails (PII detection, content moderation, schema constraints, and domain rules), and returns the best-scoring output with selection metadata. The available operational readout reports \num{5} attempts within \SI{20}{s} and \SI{91}{\percent} compliance.

For payments dispute defense summaries, we analyze aggregate operational scenario readouts rather than a randomized A/B test. Variable cohorts show higher count win rates than controls overall (\num{301}/\num{659} vs. \num{536}/\num{1548}; +\num{11.0} pp, 95\% CI [\num{6.6}, \num{15.5}], $p<0.001$) and for adjusted item-not-received cases (+\num{7.5} pp, 95\% CI [\num{0.2}, \num{15.7}], $p=0.045$). Fraud and local evidence-ranking deltas are directionally positive but not statistically significant from the aggregate count data. We also report reviewer-calibrated Responsible-AI evidence-quality signals from \num{770} generated-evidence reviews and a \num{70}-case OCR slice, and we document the reproducibility boundary through the request interface, scoring logic, pseudocode, and operational evidence boundary.
\end{abstract}


\section{Introduction}
Enterprises increasingly rely on automated document generation for compliance, finance, and customer operations. Unlike open-ended text generation, operational documents are typically \emph{schema-constrained} (JSON outputs, fixed sections, strict length caps), \emph{policy-bounded} (forbidden terms, required inclusions, privacy constraints), and \emph{high-volume} (tight latency and throughput requirements). Large language models (LLMs) provide strong fluency, but production deployments surface reliability and safety risks: schema violations, missing required fields, policy breaches, and leakage of personally identifiable information (PII).

A common pre-deployment pattern is to bolt on independent gates---PII detection, content moderation, format validators---as separate services. While effective in isolation, this fragmentation creates practical issues for distributed systems: sequential latency accumulation \cite{b_dean2013}, inconsistent rule interpretation across products, duplicated engineering effort, and higher manual rework when failures require human triage.

This paper focuses on \emph{guardrails as a first-class distributed systems component}. We introduce a unified guardrail layer that (i) standardizes policy enforcement across use cases, (ii) supports \emph{multimodal} inputs (text and images), and (iii) uses an explicit compliance score to drive \emph{best-of-$N$} selection and early exit.

\subsection{Contributions}
\begin{itemize}
\item \textbf{Unified guardrail orchestration:} a configuration-first framework that ``defines once and enforces everywhere,'' consolidating PII, moderation, schema checks, and domain rules.
\item \textbf{Compliance-scored best-of-$N$:} a multi-head generation and scoring loop that returns the highest-compliance candidate, stopping early at a configurable threshold.
\item \textbf{Multimodal support:} a single guardrail surface for text and image evidence (e.g., proof-of-delivery screenshots), enabling consistent privacy and safety enforcement.
\item \textbf{Transparent operational evaluation:} aggregate scenario-level outcome comparisons with explicit non-randomized design caveats, count-based confidence intervals, effect sizes, and Responsible-AI review signals.
\end{itemize}

\section{Related Work}
\label{sec:related}
\textbf{Safety, moderation, and privacy for LLMs.} Safety work in LLM deployment spans model-level alignment, runtime moderation, and system-level governance. Instruction tuning and RLHF reduce harmful behavior at the model layer \cite{b_ouyang2022}, while Constitutional AI replaces some human preference labels with rule-based critique \cite{b_bai2022}. Recent moderation-specific systems such as Llama Guard and Llama Guard 3 provide input/output safeguard models and compact deployment variants \cite{b_inan2023,b_fedorov2024}. Other recent guardrail models focus on adversarial robustness and explicit reasoning over safety categories, including RigorLLM and $R^2$-Guard \cite{b_yuan2024,b_kang2024}. These approaches are complementary to the runtime orchestration proposed here: they can be used as individual guardrail checks inside the scoring service, but do not by themselves define a full production request path with candidate selection, metadata, and fallback routing.

\textbf{Programmable and policy-level guardrails.} System-level governance practices such as model cards emphasize documenting intended use, limitations, and monitoring \cite{b_mitchell2019}; recent risk frameworks and practitioner taxonomies emphasize prompt injection, insecure output handling, data leakage, and human oversight for LLM applications \cite{b_nist2024,b_owasp2025}. NeMo Guardrails shows how programmable rails can control conversational LLM applications through interpretable runtime policies \cite{b_rebedea2023}. Our work extends this direction to multimodal structured document generation, where the policy engine must combine PII checks, moderation, schema validation, domain rules, early-exit selection, and asynchronous observability.

\textbf{Prompt injection and policy bypass.} Prompt injection highlights the brittleness of relying solely on prompt text to enforce constraints. Attacks can cause models to ignore developer instructions or reveal sensitive context \cite{b_perez2022}. The OWASP LLM taxonomy treats prompt injection as a first-order application security risk \cite{b_owasp2025}. More recent work also shows that guardrail judgments can shift under retrieval-augmented contexts, motivating explicit monitoring of the context in which a guardrail is evaluated \cite{b_she2025}. This motivates guardrails that are evaluated after generation as an external check, and in some cases enforced during generation via constrained decoding.

\textbf{Structured and constrained generation.} Many enterprise use cases require strict schemas (JSON, fixed fields). Constrained decoding and grammar-based generation provide mechanisms to guarantee structure \cite{b_hokamp2017}. Recent guided-generation work reformulates constrained neural decoding through finite-state machines and grammar constraints, improving reliability for structured interfaces \cite{b_willard2023}. In practice, production systems often combine constrained formats with validation and repair loops that re-prompt or regenerate when failures occur.

\textbf{Best-of-$N$ and self-consistency.} Multi-sample decoding and selection is a practical inference-time strategy for improving quality without retraining. Self-consistency demonstrates that sampling multiple reasoning traces and aggregating can improve correctness \cite{b_wang2022,b_wei2022}. We apply a related principle to compliance: sample multiple candidates and select the one maximizing an explicit guardrail score.

\textbf{Retrieval-augmented generation and evidence grounding.} Retrieval-augmented generation (RAG) conditions generation on external evidence to reduce hallucinations \cite{b_lewis2020}. Payments dispute packages similarly rely on evidence such as tracking scans, AVS/CVV signals, proof-of-delivery images, and customer/seller message context. Guardrails complement RAG by enforcing evidence inclusion, banning unverifiable claims, and routing risky outputs for review.

\textbf{Evaluation and operational ML.} Holistic evaluation frameworks such as HELM and VHELM emphasize that deployed language and vision-language systems must be measured across dimensions such as robustness, toxicity, fairness, safety, and efficiency rather than accuracy alone \cite{b_liang2022,b_lee2024}. PROMPTEVALS further highlights the practical need for assertions and guardrails in production LLM pipelines \cite{b_vir2025}. From the systems perspective, production ML systems incur hidden technical debt through glue code, pipeline complexity, and monitoring gaps \cite{b_sculley2015}, while tail latency dominates user experience at scale \cite{b_dean2013}. The guardrail layer is designed as a distributed component that keeps request-path enforcement synchronous while pushing telemetry aggregation and configuration updates off the critical path.

\section{Legacy Baseline: Fragmented Guardrail Gates}
\label{sec:baseline}
Before the unified guardrail layer, production document-generation flows resembled a pipeline of independent gates: an LLM produced a draft, then PII detection and content moderation were applied as separate services; failures triggered manual triage or re-prompting. This pattern has two core drawbacks.

\textbf{(1) Latency and retry amplification.} Each additional synchronous service call adds tail latency in the request path \cite{b_dean2013}. When a gate fails, the system either blocks the request for manual review or initiates retries/re-prompts, further increasing latency and cost.

\textbf{(2) Inconsistent enforcement and operational overhead.} Different teams implement ``similar'' rules (e.g., what constitutes PII leakage) in slightly different ways, leading to divergence over time. Fixes must be replicated across services, and monitoring becomes fragmented.

\begin{figure}[!t]
\centering
\includegraphics[width=\linewidth]{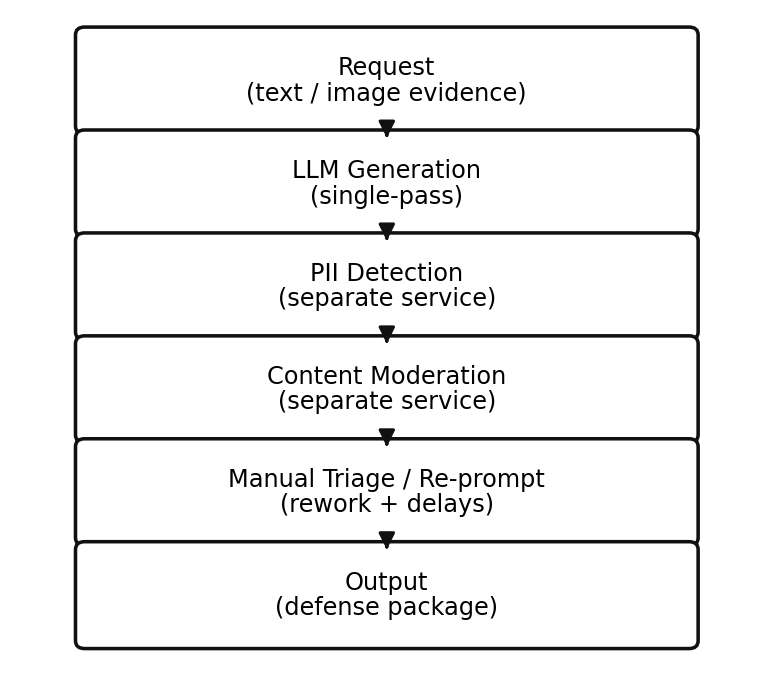}
\caption{Legacy baseline pipeline. A single-pass generation step is followed by separate PII detection, moderation, and manual triage/re-prompting before a defense package is emitted.}
\label{fig:legacy-pipeline}
\end{figure}

\begin{figure}[!t]
\centering
\includegraphics[width=\linewidth]{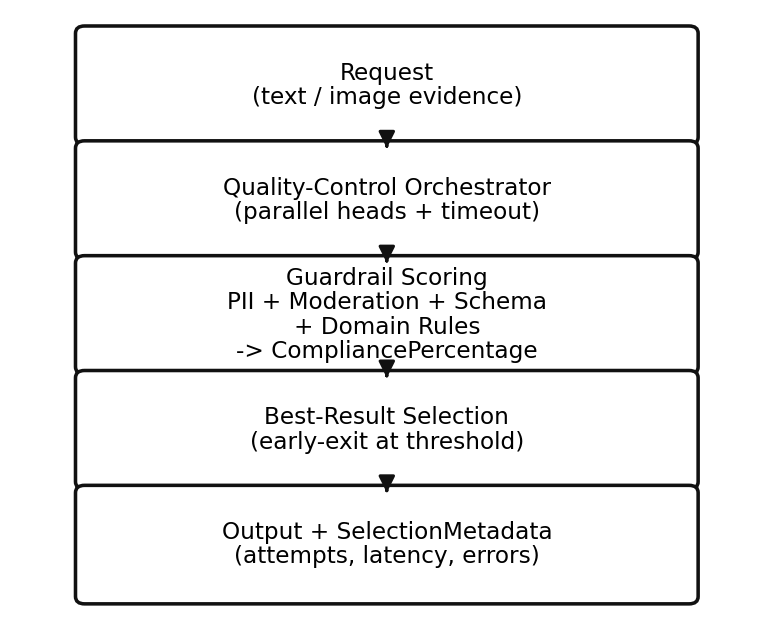}
\caption{Unified guardrail-driven pipeline. Candidate generation, guardrail scoring, best-result selection, and selection metadata are integrated into the request path.}
\label{fig:unified-pipeline}
\end{figure}

Figures~\ref{fig:legacy-pipeline} and~\ref{fig:unified-pipeline} illustrate the architectural difference between the pre-guardrail baseline and a unified guardrail-driven pipeline.

\begin{table}[!t]
\caption{Qualitative comparison of baseline vs. unified guardrail layer.}
\label{tab:compare}
\centering
\begin{tabular}{p{0.46\linewidth}cc}
\toprule
\textbf{Capability} & \textbf{Baseline} & \textbf{Unified layer} \\
\midrule
Single enforcement surface for policies & \texttimes & \checkmark \\
Integrated PII + moderation + schema & \texttimes & \checkmark \\
Best-of-$N$ selection with early exit & \texttimes & \checkmark \\
Selection metadata (attempts/latency/errors) & \texttimes & \checkmark \\
Config-first rule iteration (no code changes) & \texttimes & \checkmark \\
Multimodal evidence (text + image) & partial & \checkmark \\
\bottomrule
\end{tabular}
\end{table}

\section{Guardrail Orchestration Framework}
\label{sec:framework}
\begin{figure}[!t]
\centering
\includegraphics[width=\linewidth]{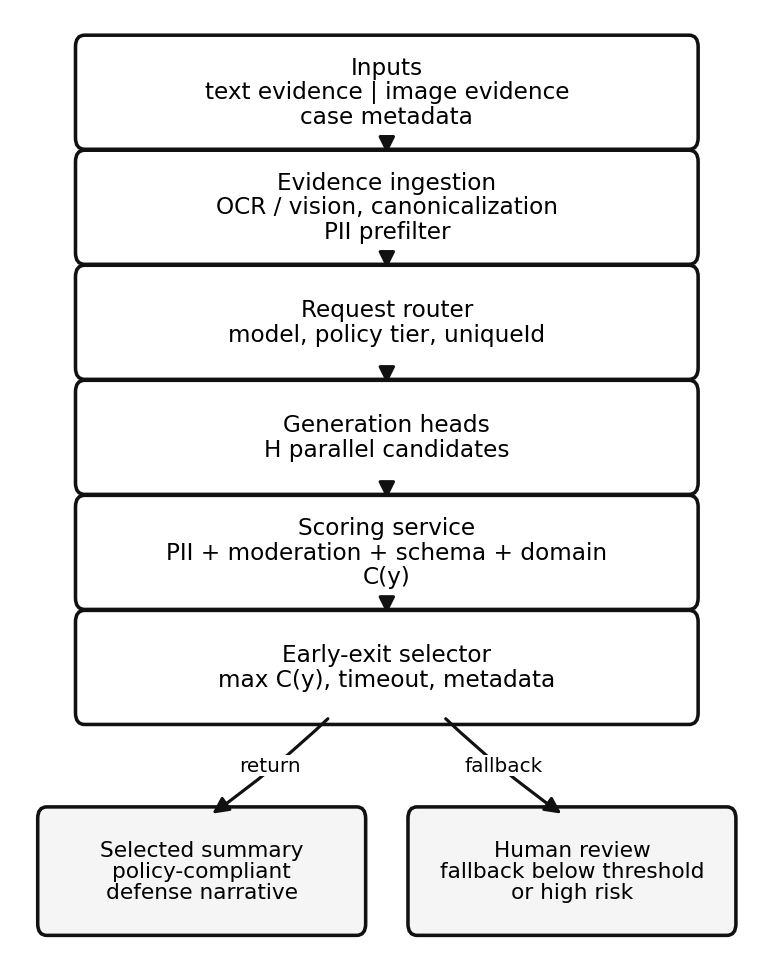}
\caption{Synchronous request path. Evidence is ingested and canonicalized; the request router selects the model and policy tier; generation heads produce candidates; the scoring service evaluates PII, moderation, schema, and domain rules; the early-exit selector returns a compliant summary or routes below-threshold/high-risk cases to human review.}
\label{fig:request-path}
\end{figure}

\begin{figure}[!t]
\centering
\includegraphics[width=\linewidth]{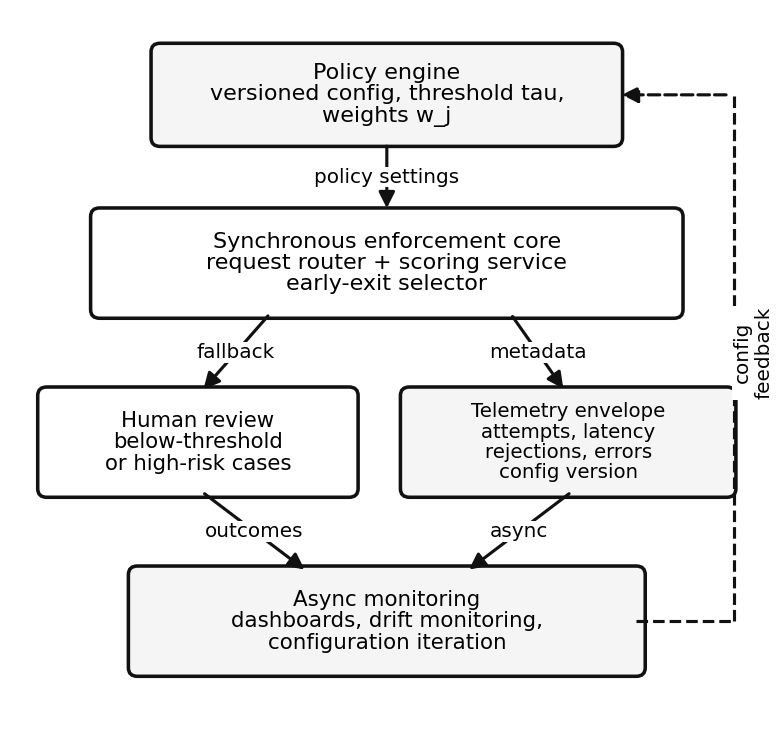}
\caption{Control and observability loop. Policy settings remain on the synchronous path, while telemetry, dashboards, drift monitoring, review outcomes, and configuration feedback are asynchronous.}
\label{fig:control-loop}
\end{figure}

The guardrail layer is designed as a reusable library and service that embeds a guardrail-driven selection loop into distributed generation systems. Key design principles are: \emph{config-first policy}, \emph{parallel candidate exploration}, \emph{explicit compliance scoring}, and \emph{operational telemetry}. Figures~\ref{fig:request-path} and~\ref{fig:control-loop} split the design into the synchronous request path and the asynchronous monitoring/configuration loop.

\subsection{Guardrail Set and Compliance Score}
We represent guardrails as a collection of checks $\mathcal{G} = \{g_1,\ldots,g_m\}$, where each $g_j(y)\in[0,1]$ measures whether candidate output $y$ satisfies a constraint (e.g., schema validity, disallowed terms, PII leakage, toxicity). Each check has a nonnegative weight $w_j$.

\begin{equation}
\begin{aligned}
C(y) &= \frac{\sum_{j=1}^{m} w_j\, g_j(y)}{\sum_{j=1}^{m} w_j} \in [0,1],\\
\text{CompliancePercentage}(y) &= 100\cdot C(y).
\end{aligned}
\label{eq:compliance}
\end{equation}

This normalization enables uniform thresholding across heterogeneous rule sets. Domain rules can be included as guardrails (e.g., ``do not mention prohibited dispute terms'' or ``ensure a boolean is returned'') and weighted by severity.

\subsection{Best-of-$N$ Candidate Generation with Early Exit}
The core runtime loop is ``best-of-$N$ under a latency budget.'' For each request, the orchestrator generates candidates in parallel and scores each against Eq.~\eqref{eq:compliance}. If a candidate meets a compliance threshold $\tau$, the system exits early.

\begin{figure}[!t]
\centering
\includegraphics[width=\linewidth]{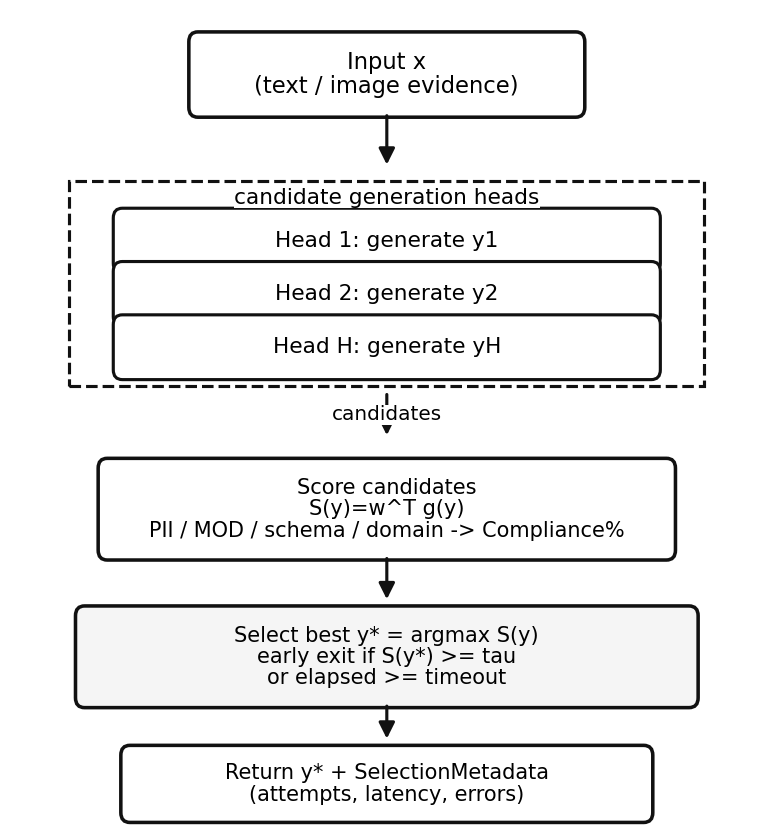}
\caption{Multi-head best-of-$N$ quality control with explicit compliance scoring and early exit at threshold $\tau$.}
\label{fig:qc}
\end{figure}

Figure~\ref{fig:qc} illustrates the quality-control loop, and Algorithm~\ref{alg:guardrail} summarizes the core logic.

\begin{algorithm}[!t]
\caption{Guardrail-Driven Best-of-$N$ Generation}
\label{alg:guardrail}
\begin{algorithmic}[1]
\REQUIRE input $x$ (text/image evidence), guardrails $\mathcal{G}$, weights $w$, threshold $\tau$, timeout $T_{\max}$, heads $H$
\STATE $y^*\leftarrow \varnothing$, $c^*\leftarrow -\infty$, start timer
\WHILE{elapsed $<T_{\max}$}
\STATE Generate $H$ candidates in parallel: $\{y_1,\ldots,y_H\} \leftarrow \textsc{Generate}(x)$
\FOR{each $y_h$}
\STATE $c_h \leftarrow C(y_h)$ using Eq.~\eqref{eq:compliance}
\IF{$c_h > c^*$} \STATE $y^*\leftarrow y_h$, $c^*\leftarrow c_h$ \ENDIF
\ENDFOR
\IF{$c^* \ge \tau$} \STATE \textbf{break} \ENDIF
\ENDWHILE
\STATE return $y^*$ and SelectionMetadata (attempts, latency, errors)
\end{algorithmic}
\end{algorithm}

Algorithm~\ref{alg:normalize} makes the multimodal enforcement surface explicit: both text and image evidence are normalized into a single representation before scoring, ensuring that privacy and safety guardrails apply consistently across modalities.

\begin{algorithm}[!t]
\caption{Multimodal Evidence Normalization and Privacy Filtering}
\label{alg:normalize}
\begin{algorithmic}[1]
\REQUIRE text evidence $e^{\text{text}}$, optional image evidence $e^{\text{img}}$, flags (PII, moderation), redaction policy $\rho$
\STATE $u \leftarrow e^{\text{text}}$
\IF{$e^{\text{img}}$ present}
  \STATE $u \leftarrow u \oplus \textsc{ExtractText}(e^{\text{img}})$ \COMMENT{OCR or vision-capable model}
\ENDIF
\STATE $u \leftarrow \textsc{Canonicalize}(u)$ \COMMENT{normalize whitespace, timestamps, locales}
\IF{PII enabled}
  \STATE $(p, \lambda) \leftarrow \textsc{DetectPII}(u)$ \COMMENT{entities + confidence}
  \STATE $u \leftarrow \textsc{Redact}(u, p, \rho)$ \COMMENT{mask identifiers, preserve semantics}
\ENDIF
\IF{Moderation enabled}
  \STATE $m \leftarrow \textsc{Moderate}(u)$
  \IF{$m$ violates policy} \STATE flag violation (score penalty) \ENDIF
\ENDIF
\STATE return normalized evidence $u$ and enforcement flags
\end{algorithmic}
\end{algorithm}

Algorithm~\ref{alg:async} highlights the production deployment detail that is easy to miss in purely offline evaluations: enforcement and scoring must be synchronous to guarantee policy compliance, while telemetry, dashboards, and configuration updates are pushed off the critical path.

\begin{algorithm}[!t]
\caption{Synchronous Enforcement with Asynchronous Monitoring and Policy Iteration}
\label{alg:async}
\begin{algorithmic}[1]
\REQUIRE request $x$, cached guardrail config $\theta$ (versioned), async queue $Q$
\STATE $\theta \leftarrow \textsc{GetConfig}()$ \COMMENT{fast path: cached; slow path: refresh}
\STATE $x' \leftarrow \textsc{NormalizeEvidence}(x)$ using Alg.~\ref{alg:normalize}
\STATE $(y^*, c^*, meta) \leftarrow \textsc{BestOfN}(x',\theta)$ using Alg.~\ref{alg:guardrail}
\STATE \textbf{return} $y^*$ \COMMENT{synchronous: only policy-compliant output is returned}
\STATE enqueue $(meta, c^*, \theta\_\text{version})$ to $Q$ \COMMENT{asynchronous}
\STATE background worker: aggregate metrics; update dashboards; propose $\theta$ changes; hot-reload
\end{algorithmic}
\end{algorithm}

\textbf{Synchronous vs asynchronous updates.} Validation and compliance scoring occur synchronously within the request path to guarantee that returned outputs meet policy thresholds. Telemetry aggregation, dashboards, and configuration updates are applied asynchronously to avoid request-path latency and support safe iteration.

\subsection{Integrated PII and Content Moderation}
The unified guardrail layer treats privacy and safety checks as first-class guardrails rather than external gates. In addition to format checks (e.g., strict JSON schema) and domain-specific rules (e.g., required evidence inclusion), the compliance score includes:
\begin{itemize}
\item \textbf{PII detection:} penalizes outputs that expose sensitive identifiers (names, addresses, account numbers, tracking identifiers), optionally applying redaction.
\item \textbf{Content moderation:} filters policy-violating or disallowed content using a configurable severity tier.
\end{itemize}
\FloatBarrier

\subsection{Multimodal Inputs: Images and Text}
Payments disputes frequently include images (shipping labels, proof-of-delivery screenshots). The unified guardrail layer supports multimodal evidence by (i) extracting text via OCR or a vision-capable model, and (ii) applying the same PII and moderation guardrails across both modalities. This unifies enforcement regardless of evidence type and reduces the need for ad hoc image-handling workflows.

\section{Integration Surface}
\label{sec:api}
The guardrail layer is exposed as a service endpoint that accepts evidence and policy knobs as request parameters. The following fields (derived from a production-style request invocation) illustrate the integration surface:

\begin{table}[!t]
\caption{Example guardrail API surface (request fields).}
\label{tab:api}
\centering
\begin{tabular}{p{0.26\linewidth} p{0.67\linewidth}}
\toprule
\textbf{Field} & \textbf{Purpose} \\
\midrule
\texttt{text} & Evidence text (e.g., tracking events) \\
\texttt{image} & Optional image evidence (multipart) \\
\texttt{modelName} & Target model (sandbox vs production) \\
\texttt{piiDetection} & Enable/disable PII checking and redaction \\
\texttt{Content \allowbreak moderation} & Enable/disable moderation checks \\
\texttt{moderation\allowbreak Level} & Moderation strictness (policy tier) \\
\texttt{guardrail} & Domain rule (e.g., ``ensure boolean returned'') \\
\texttt{guardrail\allowbreak Level} & Domain strictness / threshold knob \\
\texttt{uniqueId} & Request identifier for audit and traceability \\
\bottomrule
\end{tabular}
\end{table}

\section{Evaluation and Results}
\label{sec:results}
We evaluate the unified guardrail layer along four axes: (i) evaluation design and statistical assumptions, (ii) guardrail effectiveness and operational characteristics, (iii) Responsible-AI evidence-quality review signals, and (iv) downstream outcome impact in payments dispute defense workloads.

\subsection{Evaluation Design and Statistical Method}
The payments outcome readouts in Table~\ref{tab:outcomes} are \emph{operational scenario comparisons}, not a randomized A/B test. The variable groups correspond to guardrail-enabled AI defense-summary or evidence-ranking scenarios; controls correspond to operational control cohorts from the same reporting context. Because the available exports do not prove random assignment, blocking, or matched-pair construction, we interpret the observed differences as \emph{associations} and avoid claiming that the guardrail system alone caused the win-rate lift. Residual confounding is possible from case mix, dispute reason, network policy, seller behavior, evidence availability, and time-window effects.

For count-based win rates, we recover approximate wins by multiplying the reported dispute counts by the reported percentages and rounding to the nearest integer. We then compute Wilson 95\% confidence intervals for each rate, Newcombe/Wilson intervals for rate differences, two-proportion z-tests, relative risk (RR), and odds-ratio effect sizes. This inference is therefore appropriate for aggregate count readouts, but it should be replaced by exact item-level analyses before camera-ready submission. Amount-weighted win rates are reported as effect sizes only: the PDFs provide aggregate amount-weighted percentages but not the per-dispute dollar amounts needed for valid bootstrap, permutation, or design-based confidence intervals.

\subsection{Compliance, Latency, and Public Operational Readout}
\begin{figure}[!t]
\centering
\includegraphics[width=\linewidth]{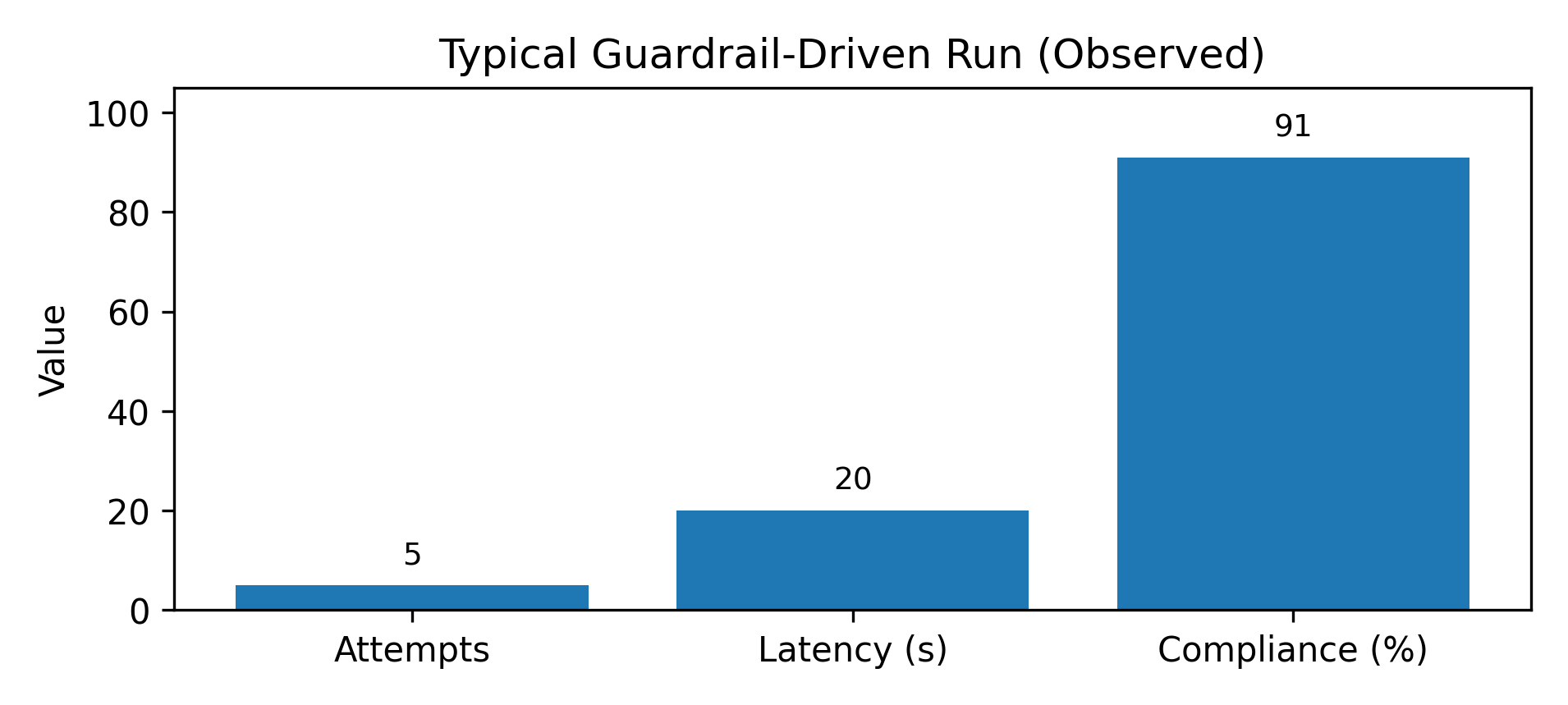}
\caption{Aggregate run characteristics disclosed in the approved public readout: candidate attempts, latency budget, and compliance. Request-level latency percentiles, cost, fallback rate, human-review reduction, and ablation metrics are not reported in this version because they require validated request-level telemetry exports.}
\label{fig:typical}
\end{figure}

The guardrail layer emits request metadata including candidate attempts, latency, compliance score, rejection reasons, fallback decisions, and configuration version. In this public version, we report only the aggregate operational readout approved for disclosure: a typical run uses \num{5} candidate attempts, completes within a \SI{20}{s} request budget, and reaches \SI{91}{\percent} compliance. Figure~\ref{fig:typical} summarizes these disclosed values.

\subsection{Responsible-AI Evidence-Quality Review Signals}
Two additional review decks provide evaluator-facing quality signals for the evidence-generation and OCR components. For generated M2M dispute evidence, the corpus contains \num{770} unique reviews, with \num{318} labeled reviews available for estimation across five reviewers. Projected to the full corpus, approximately \num{574}/\num{770} reviews are rated \num{3}/\num{5}, \num{157}/\num{770} are rated \num{4}--\num{5}, and \num{39}/\num{770} are rated \num{1}--\num{2}; the projected mean rating is \num{3.17}. Narrative reviewer detail appears in \num{74.5}\% of labeled exports, projecting to about \num{574}/\num{770} reviews, while only about \num{77}/\num{770} contain any structured issue flag. The most common structured issues are title-not-aligned (about \num{29}) and title-inaccurate (about \num{27}), indicating that a flag-only dashboard would miss much of the review nuance.

For OCR evidence quality, the slice contains \num{70} cases, five analysts, and \num{350} ratings. The fields include text accuracy, word/line integrity, noise vs. non-text confusion, overall OCR, evidence utility, and pass/fail. The global weighted median of per-case mean Overall OCR scores is \num{3.2}. Pairwise ordinal agreement is moderate to strong: the highest reported weighted kappa alignments are \num{0.753}, \num{0.741}, and \num{0.722}; disagreement concentrates in blur drop-out (mean case SD \num{0.91}), cropped text loss (\num{0.90}), and UI bleed or initial-capture artifacts (\num{0.78}). A clean chat case achieves Overall OCR \num{4.40} $\pm$ \num{0.55} with pass rate \num{1.00}, while an analyzer-error label/receipt case scores \num{1.00} $\pm$ \num{0.00} with pass rate \num{0.00}. These signals motivate a human-review fallback path for low-quality visual evidence and analyzer failures.

\subsection{Payments Dispute Defense Summaries: Outcome Readouts}
Chargeback defense summaries must be concise, evidence-backed narratives that comply with payment-network rules while avoiding prohibited language and PII leakage. Table~\ref{tab:outcomes} reports scenario-level outcomes comparing AI-driven variable scenarios with control baselines.

\begin{table}[!t]
\caption{Downstream outcome readouts (variable vs. control). ``Win\%'' is by count; ``WinAmt\%'' is amount-weighted.}
\label{tab:outcomes}
\centering
\scriptsize
\begin{tabular}{l l c c c}
\toprule
\textbf{Scenario} & \textbf{Group} & \textbf{Disp. Cnt} & \textbf{Win\%} & \textbf{WinAmt\%} \\
\midrule
Overall AI Defense Summary & Variable & 659 & 45.7 & 41.4 \\
 & Control & 1548 & 34.6 & 22.3 \\
\midrule
Fraud Defense Summary & Variable & 501 & 50.5 & 45.3 \\
 & Control & 708 & 48.7 & 38.1 \\
\midrule
INR Defense Summary (adj.) & Variable & 152 & 30.3 & 26.9 \\
 & Control & 840 & 22.7 & 19.3 \\
\midrule
Evidence Ranking (Local) & Variable & 445 & 17.30 & 24.5 \\
 & Control & 406 & 14.53 & 17.9 \\
\bottomrule
\end{tabular}
\end{table}

\begin{table}[!t]
\caption{Count inference and amount-weighted effect sizes. Count inference uses rounded wins from aggregate percentages; amount CIs/p-values require item-level disputed amounts.}
\label{tab:stats}
\centering
\resizebox{\linewidth}{!}{%
\begin{tabular}{lccccc}
\toprule
\textbf{Scenario} & \textbf{Var} & \textbf{Ctl} & \textbf{Count delta [95\% CI]} & \textbf{$p$} & \textbf{Amt / rel.} \\
\midrule
Overall AI Summary & 301/659 & 536/1548 & +11.0 [6.6, 15.5] & $<.001$ & +19.1 / 1.86x \\
Fraud Summary & 253/501 & 345/708 & +1.8 [-3.9, 7.5] & .544 & +7.2 / 1.19x \\
INR Summary (adj.) & 46/152 & 191/840 & +7.5 [0.2, 15.7] & .045 & +7.6 / 1.39x \\
Evidence Ranking & 77/445 & 59/406 & +2.8 [-2.2, 7.7] & .270 & +6.6 / 1.37x \\
\bottomrule
\end{tabular}%
}
\end{table}

\subsection{Evidence Ranking and Summary Generation Lifts}
\begin{figure}[!t]
\centering
\includegraphics[width=\linewidth]{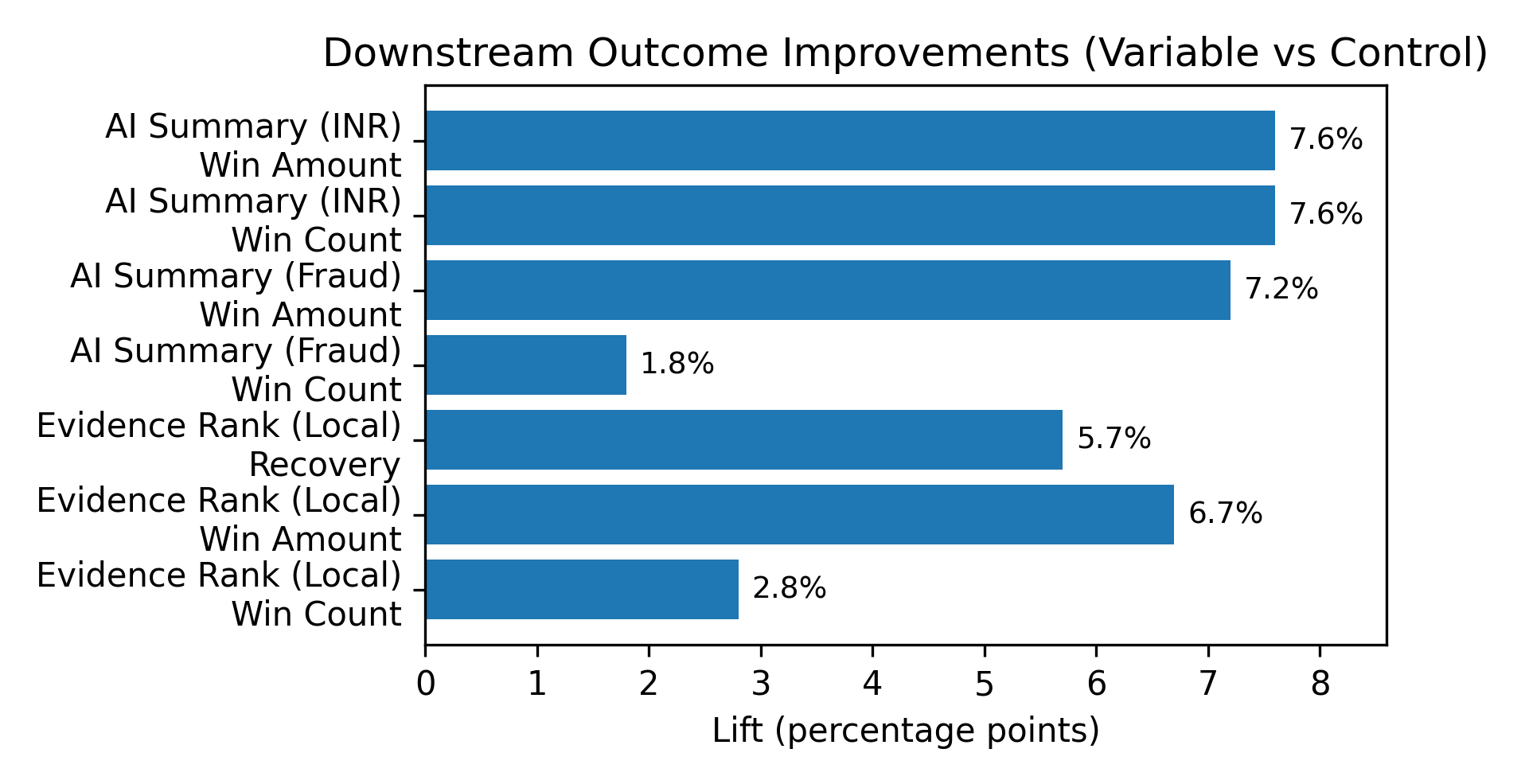}
\caption{Downstream outcome improvements (variable vs. control) in evidence ranking and AI defense-summary scenarios. These are operational cohort deltas, not randomized causal estimates.}
\label{fig:lifts}
\end{figure}

Local evidence ranking changes show a +\num{2.8} pp count win-rate delta and +\num{6.6} pp amount-weighted delta, but the count-based aggregate comparison is not statistically significant at the 0.05 level. AI-generated defense-summary scenarios show the largest overall count delta (+\num{11.0} pp) and amount-weighted delta (+\num{19.1} pp). Fraud-specific deltas are positive but not statistically significant by count; adjusted item-not-received (INR) comparisons show a +\num{7.5} pp count delta with a confidence interval barely above zero. Figure~\ref{fig:lifts} visualizes the reported outcome deltas.

\subsection{Evidence Boundary for Runtime Ablations}
The public aggregate readout supports the best-of-$N$ operating point described above, but it does not contain enough request-level records to report validated ablations for single-candidate generation, $N$ sensitivity, early-exit threshold sweeps, guardrail-weight perturbations, or compliance/outcome calibration. We therefore do not report latency percentiles, cost, fallback rate, human-review reduction, or ablation results in this version. Those analyses require request-level telemetry with stable identifiers, timestamps, model/token usage, guardrail-failure labels, reviewer actions, and downstream outcomes. Section~\ref{sec:limitations} treats this as a limitation rather than filling the missing fields with unverified values.

\section{Discussion}
The unified guardrail layer demonstrates a practical pattern for high-stakes generation systems: treat policy and format constraints as a composable set of guardrails, compute an explicit compliance score used for early exit, and allocate generation budget to maximize compliance under latency limits. This buffers from post-hoc filtering in two ways. First, selection is \emph{score-driven}: the engine returns the best candidate observed under a budget rather than accepting/rejecting a single draft. Second, policies become \emph{configuration}: changes to rules and thresholds do not require code changes, enabling faster iteration.

\subsection{Latency, Throughput, and Cost Trade-offs}
Best-of-$N$ generation increases model calls but can reduce total cost by lowering manual triage and avoiding downstream failures. A simple expected-cost model is
\begin{equation}
\mathbb{E}[\text{Cost}] = c\,\mathbb{E}[N_{\text{calls}}] + m\,\mathbb{E}[N_{\text{manual}}],
\end{equation}
where $c$ is per-call model cost and $m$ is per-item manual handling cost. The unified guardrail layer reduces $\mathbb{E}[N_{\text{manual}}]$ by enforcing PII, moderation, and schema constraints before output is emitted, while bounding $\mathbb{E}[N_{\text{calls}}]$ through early exit at threshold $\tau$.

From a systems perspective, a key advantage is that parallel heads trade \emph{compute} for lower \emph{tail risk}: under fixed budgets, selecting the maximum-compliance candidate reduces the probability of returning a non-compliant output.

\subsection{Reproducibility Boundary}
Raw dispute records, evidence screenshots, customer messages, and production code cannot be released. To support review without exposing proprietary data, the paper specifies the request interface, compliance-score formula, evidence-normalization algorithm, best-of-$N$ selection loop, and asynchronous telemetry pattern. A de-identified artifact package may be released separately after organizational approval; until then, this version should be read as an applied systems paper with public algorithms and aggregate outcome readouts, not as a fully reproducible public benchmark.

\section{Limitations}
\label{sec:limitations}
This work has several important limitations. First, the downstream outcome readouts are aggregate, non-randomized operational comparisons. Causal interpretation would require a randomized experiment, a matched cohort design, or covariate-adjusted item-level analysis. We therefore describe observed deltas as associations rather than claiming that the guardrail layer alone caused the win-rate differences.

Second, the public version does not include raw proprietary payments data, evidence screenshots, customer messages, or production code. External reproducibility is therefore limited to the algorithmic design, scoring equation, pseudocode, and disclosed aggregate metrics.

Third, amount-weighted win rates are reported as aggregate effect sizes only. Statistically valid confidence intervals for amount-weighted outcomes require per-dispute dollar amounts and an appropriate bootstrap, permutation, or design-based test.

Fourth, this version does not report request-level latency percentiles, per-summary cost, fallback rate, human-review reduction, or ablation results. Those values require validated production telemetry exports that were not available in approved aggregate form for this manuscript version.

Fifth, the system is domain-specific to payments dispute defense. The guardrail scoring pattern may transfer to other structured document-generation settings, but rule weights, thresholds, risk routing, and human-review policies must be recalibrated for each domain. The Responsible-AI review decks also identify operational risks requiring human oversight, including title drift in generated evidence, analyzer/OCR failure, cropped text loss, UI bleed, and poor handwritten/blurry-image extraction.

Future work includes contextual routing to select models/guardrail settings based on case features \cite{b_li2010}, learned calibration of weights $w_j$ in Eq.~\eqref{eq:compliance}, request-level ablations for $N$, $\tau$, and guardrail weights, and counterfactual evaluation for policy changes \cite{b_koh2021}. Another direction is safety-aware reward modeling that explicitly penalizes privacy and toxicity risks \cite{b_mitchell2019,b_bender2021}.

\section{Conclusion}
We presented a compliance-scored, best-of-$N$ guardrail orchestration framework that unifies PII detection, content moderation, schema validation, and domain rules into an explicit compliance score used for early exit. In payments dispute defense-summary scenarios, the available aggregate readouts show high compliance at bounded latency and positive operational outcome associations, with statistically significant count-based deltas overall and for adjusted INR cases. The next publication-ready step is to fill the request-level ablation, percentile latency, cost, fallback, and amount-weighted inference tables from production telemetry while retaining the human-review and reproducibility limits described above.


\begin{thebibliography}{99}

\bibitem{b_dean2013}
J.~Dean and L.~A. Barroso,
``The tail at scale,''
\emph{Communications of the ACM}, vol.~56, no.~2, pp.~74--80, 2013.

\bibitem{b_sculley2015}
D.~Sculley et al.,
``Hidden technical debt in machine learning systems,''
in \emph{Proc. NeurIPS (Workshop)}, 2015.

\bibitem{b_ouyang2022}
L.~Ouyang et al.,
``Training language models to follow instructions with human feedback,''
in \emph{Proc. NeurIPS}, 2022.

\bibitem{b_bai2022}
Y.~Bai et al.,
``Constitutional AI: Harmlessness from AI feedback,''
arXiv:2212.08073, 2022.

\bibitem{b_mitchell2019}
M.~Mitchell et al.,
``Model cards for model reporting,''
in \emph{Proc. FAT*}, 2019.

\bibitem{b_bender2021}
E.~M. Bender et al.,
``On the dangers of stochastic parrots: Can language models be too big?''
in \emph{Proc. FAccT}, 2021.

\bibitem{b_perez2022}
E.~Perez and M.~Ribeiro,
``Ignore previous prompt: Attack techniques for language models,''
arXiv:2211.09527, 2022.

\bibitem{b_hokamp2017}
C.~Hokamp and Q.~Liu,
``Lexically constrained decoding for sequence generation using grid beam search,''
in \emph{Proc. ACL}, 2017.

\bibitem{b_wang2022}
X.~Wang et al.,
``Self-consistency improves chain of thought reasoning in language models,''
arXiv:2203.11171, 2022.

\bibitem{b_wei2022}
J.~Wei et al.,
``Chain-of-thought prompting elicits reasoning in large language models,''
arXiv:2201.11903, 2022.

\bibitem{b_lewis2020}
P.~Lewis et al.,
``Retrieval-augmented generation for knowledge-intensive NLP tasks,''
in \emph{Proc. NeurIPS}, 2020.

\bibitem{b_li2010}
L.~Li et al.,
``A contextual-bandit approach to personalized news article recommendation,''
in \emph{Proc. WWW}, 2010.

\bibitem{b_koh2021}
P.~W. Koh et al.,
``WILDS: A benchmark of in-the-wild distribution shifts,''
in \emph{Proc. ICML}, 2021.

\bibitem{b_rebedea2023}
T.~Rebedea, R.~Dinu, M.~Sreedhar, C.~Parisien, and J.~Cohen,
``NeMo Guardrails: A toolkit for controllable and safe LLM applications with programmable rails,''
arXiv:2310.10501, 2023.

\bibitem{b_inan2023}
H.~Inan et al.,
``Llama Guard: LLM-based input-output safeguard for human-AI conversations,''
arXiv:2312.06674, 2023.

\bibitem{b_fedorov2024}
I.~Fedorov et al.,
``Llama Guard 3-1B-INT4: Compact and efficient safeguard for human-AI conversations,''
arXiv:2411.17713, 2024.

\bibitem{b_yuan2024}
Z.~Yuan, Z.~Xiong, Y.~Zeng, N.~Yu, R.~Jia, D.~Song, and B.~Li,
``RigorLLM: Resilient guardrails for large language models against undesired content,''
arXiv:2403.13031, 2024.

\bibitem{b_kang2024}
M.~Kang and B.~Li,
``$R^2$-Guard: Robust reasoning enabled LLM guardrail via knowledge-enhanced logical reasoning,''
arXiv:2407.05557, 2024.

\bibitem{b_willard2023}
B.~T. Willard and R.~Louf,
``Efficient guided generation for large language models,''
arXiv:2307.09702, 2023.

\bibitem{b_liang2022}
P.~Liang et al.,
``Holistic evaluation of language models,''
arXiv:2211.09110, 2022.

\bibitem{b_lee2024}
T.~Lee et al.,
``VHELM: A holistic evaluation of vision language models,''
arXiv:2410.07112, 2024.

\bibitem{b_vir2025}
R.~Vir, S.~Shankar, H.~Chase, W.~Fu-Hinthorn, and A.~Parameswaran,
``PROMPTEVALS: A dataset of assertions and guardrails for custom production large language model pipelines,''
arXiv:2504.14738, 2025.

\bibitem{b_she2025}
Y.~She, D.~W. Peterson, M.~M. Liu, V.~Upadhyay, M.~H. Chaghazardi, E.~Kang, and D.~Roth,
``RAG makes guardrails unsafe? Investigating robustness of guardrails under RAG-style contexts,''
arXiv:2510.05310, 2025.

\bibitem{b_nist2024}
National Institute of Standards and Technology,
``Artificial Intelligence Risk Management Framework: Generative Artificial Intelligence Profile,''
NIST AI 600-1, 2024.

\bibitem{b_owasp2025}
OWASP Foundation,
``OWASP Top 10 for Large Language Model Applications 2025,''
2025. [Online]. Available: \url{https://owasp.org/www-project-top-10-for-large-language-model-applications/}

\end{thebibliography}
\end{document}